\documentstyle[preprint,aps]{revtex}

\begin{document}

\preprint{DUKE-TH-97-156}

\input epsf
\draft

\title{
Multiplicity Distribution and Mechanisms of the
High-Energy Hadron Collisions}

\author{S. G. Matinyan\footnote{Also at Yerevan Physics Institute, 
Yerevan, 375036 Armenia} and W. D. Walker}

\address{Department of Physics, Duke University \\
Durham, NC 27708-0305, USA}

\date{\today}

\maketitle

\begin{abstract}

We discuss the multiplicity distribution for highest accessible
energies of $pp$- and $\bar pp$- interactions from the point of view
of the multiparton collisions.  The inelastic cross sections for the
single, $\sigma_1$,  and multiple (double and, presumably, triple,
$\sigma_{2+3}$) parton collisions are extracted from the analysis of
the experimental data on the multiplicity distribution up to the
Tevatron energies.  It follows that $\sigma_1$ becomes energy
independent while $\sigma_{2+3}$ increases with $\sqrt{s}$ for
$\sqrt{s}\ge$ 200 GeV.  The observed growth of $\langle
p_{\perp}\rangle$ with multiplicity is attributed to the increasing
role of multiparton collisions for the high energy $\bar pp(pp)$-
inelastic interactions.

\end{abstract}


\section{Introduction.  KNO scaling and its Violation}

The experimental observations of the violation of KNO scaling
\cite{ZK72} at the high energy $(pp)$- and $(\bar pp)$-collisions
\cite{UA5} and of the correlations between the average transverse
momentum $\langle p_{\perp}\rangle$ and the multiplicity $N$ of the
secondaries (a higher $\langle p_{\perp}\rangle$ for high multiplicity
events) \cite{UA1} indicate that there are at least two
(complementary) mechanisms of the high energy multiparticle production
revealing the quark-gluon structure of hadrons and their interactions.

That KNO scaling should be violated at very high energies was realized 
long ago \cite{VAA72}, in the same year when KNO scaling was introduced.  
This deviation from a KNO type distribution of secondaries was attributed 
to the possibility of the splitting of each of the colliding hadrons into 
several constituents (valence quarks, partons) pair wise interacting with 
their counterparts from oppositely moving hadrons.  The above picture 
results in the production of several showers and, in the frame of the
Reggeism, is attributed to the contribution of Regge cuts to the elastic 
scattering amplitude, in addition to the Regge pole (Pomeron) 
corresponding to the production of the single shower.

Inclusion of such contributions changes the structure of the
distribution ${\sigma_N\over \sigma_{\rm tot}}$ $(\sigma_N$ is the
cross section of the production of $N$ secondary hadrons) at large
$\xi = \ln {s\over s_0}$ ($s_0$ = 1 GeV$^2$) leading to the appearance
of the additional peaks in the distribution with larger $\langle
N_n\rangle$ (here $n$ denotes the number of pairs of the simultaneously 
colliding partons involved in the interaction from the different hadrons, 
or the number of resulting showers; $n=1$, a single interaction, can be 
attributed to the Regge pole; $n=2$, a double collision of two pairs of 
partons from different hadrons, can be attributed to the exchange of two 
Pomerons (Regge cut), etc.).

\section{Regge Picture and Multiple Collisions of Partons}

In this picture, $\langle N_n\rangle = n \langle N_1\rangle$ where
$\langle N_1\rangle$ is the mean multiplicity corresponding to the
production of one shower in the single partonic collision.  We have,
from the data where the one Pomeron exchange describes experiments at 
lower energies $\langle N_1\rangle = a+b \ln {S\over s_0}$, with $a=
-3.879$; $b=2.099$ (see Appendix A for details).  This qualitative 
picture \cite{VAA72,KAM73} corresponds to the idealized scenario where
only ``non-enhanced'' Regge-type diagrams are included
(``quasi-eikonal'' approximation).  Inclusion of the so-called
``enhanced'' diagrams where the possibility of the branching of
ladders-showers is allowed, thus giving the emission of the smaller
average number of emitted particles than the corresponding
non-enhanced cut, leads to the smearing out of the distribution
between the peaks reducing the effects of KNO violations.  Since the
distributions of particles in the different ladders are independent,
one expects that $\Delta N_n \sim \sqrt{n N_1}$ giving the broadening
of the distribution ${\sigma_N\over \sigma_{\rm tot}}$.\footnote{Since
at the highest achievable energies $(\sqrt{s} \ge 500$ GeV) KNO
scaling (in the whole rapidity interval) is violated strongly, this
indicates that at these energies the ``enhanced'' graphs, probably,
are not important.  It is not excluded, that at higher energies KNO
scaling will be restored.}

We emphasize that the multipomeron exchanges are especially important
in the modern treatment of the Pomeron as a pole with intercept
$\alpha(0)$ higher than unity $(\alpha(0)=1+\Delta$; for ``soft'' Pomeron
$\Delta \approx 0.08$ as experimental data on high energy ``soft''
collisions show), where the relative role of these exchanges is
increased with energy \cite{PEV76} (e.g., Pomeron contribution to the
$\sigma_{\rm tot}$ is $\sim e^{\xi\Delta}$, whereas the $n$ Pomeron
exchange is $\sim e^{n\xi\Delta}$).  Concerning the effective number of 
exchanges in ``quasi-eikonal'' approximation, one has $n_{\rm eff} 
\sim e^{\xi_n\Delta} \le 2.5\left(\xi_n=\ln({s\over s_0 n^2})\right)$ 
\cite{PEV76} for $\sqrt{s} \le 2$ TeV $(n=2)$.  Thus, from this point 
of view, only double parton interactions (two effective Pomeron exchanges) 
in addition to the single collision are expected to be effective at the 
energies up to the Tevatron energy.  At LHC energy ($\sqrt{s} = 14$ TeV) 
this parameter $n_{\rm eff} \approx 3.3\; (n=3)$, thus indicating the 
possibility of the appearance of the third maximum in ${\sigma_N\over 
\sigma_{\rm tot}}$ distribution with $\langle N_3\rangle \approx 3
\langle N_1\rangle$  (see Fig. 1 and Appendix B).

\section{Gluon String Model and Pomerons}

On the QCD level, the ultra-high energy interaction between hadrons
phenomenologically described by Pomeron exchange is treated as a
result of a (few) gluons exchange between the constituents of hadrons
as they pass close to one another \cite{FEL75}.  After a color
exchange between partons (valence quarks, gluons, etc.) carried by the
colliding hadrons, partons from one hadron are joined by two gluon
strings with partons inside the other hadron.  This pair of strings
stretch when hadrons separate after the collision and break down
transforming into the system of hadrons forming a Pomeron-like
shower (ladder), giving the asymptotic density of hadrons.

From this point, the above described $n$ Pomeron exchanges are
considered as the joining and stretching of $n$ different pairs of strings 
between hadron constituents resulting, after the hadron separation, the
production of $n$ Pomeron-like showers with $n$-fold increased average
multiplicity $\langle N_n\rangle$.

This picture using the language of gluon strings can be related to the
Pomeron phenomenology, at least in the quasi-eikonal approximation,
and, as shown in \cite{ABK83}, leads to the same expressions for a
variety of characteristics of the high energy strong interactions as
the Regge diagram technique, giving also the possibility to determine
some free parameters of last one.

\section{Two Component Picture of the Hadron Collisions}

From the more phenomenological approach \cite{TH85}, it is possible to 
argue that there are at least two components in the cross sections for the
ultra-fast hadron collisions:  soft component corresponding to the
single collisions of partons from different hadrons which can (but not
necessarily) be described by the exchange of the single Pomeron (or
the pair of gluonic strings), and the so-called semi-hard (jet)
component which is responsible for the higher $\langle
p_{\perp}\rangle$ of the secondaries than is usually attributed to the
single ``soft'' Pomeron exchange.

At low energies (up to energy of ISR) only soft interactions with
scaled multiplicity distribution of the KNO type are important.  At
higher energies (CERN SPS, Tevatron) the KNO-like shape of the
multiplicity distribution is modified by the increasing role of the
``hard'' multiparton interactions \cite{TH85} (four parton collisions
of \cite{BH83}) that leads to the increase of the fraction of events
with higher multiplicity $(N \gg \langle N_1\rangle )$.  This new channel 
is characterized, at the same time, by the higher $\langle p_{\perp}
\rangle$, in accordance with the experimental observation \cite{GC85} 
that the high multiplicity events indicate a jet-like structure.

From the point of view of the interactions between constituents, one
can argue that this second ``semihard'' component with $\langle
N_2\rangle \approx 2 \langle N_1\rangle$ is due to the double collisions
when in the same inelastic interaction, two different parton pairs 
from colliding hadrons independently collide.  Since the parton densities 
of the colliding hadrons overlap strongly at small impact parameters 
(higher $p_{\perp}$) the probability of the double parton collision 
$(n=2)$ is enhanced, and, as a result, two pairs of soft partons (e.g., 
valence quarks) $(x_{\perp} = {p_{\perp}\over \sqrt{s}} \ll 1$) stacked 
in the collision resulting in the (mini) jet structure of the secondaries 
with $N \ge \langle N_2\rangle = 2\langle N_1\rangle$.

In \cite{TH85} it was shown that the above two-component model gives a
successful fit for the total non-single diffractive inelastic cross
section inclusive spectra and multiplicity distribution in full phase
space for the energies $\sqrt{s} = 200$ and 546 GeV.  We note that in
\cite{TH85} the ad hoc assumption was made that the inelastic cross
section for the soft component (single parton collisions) is energy
independent.

We will see below, from our analysis of the multiplicity distribution,
that the high energy inelastic cross section corresponding to the
single parton collisions indeed, practically does not depend on
$\sqrt{s}$, as it was assumed in \cite{TH85}.  On the other hand, the
one Pomeron exchange in its quasi-eikonal form gives an increase of
inelastic cross section.  We don't know the answer to that
inconsistency, except that the role of the enhanced diagrams is
underestimated, and they have to be included in the analysis of the
``soft'' cross section.

\section{$\langle p_{\perp}\rangle$ Versus Multiplicity}

Now, if we again turn to the language of the Pomeron exchanges, it is
possible to conjecture that the above described double parton
collisions can be attributed to the double effective Pomeron exchange which
leads not only to the higher multiplicity (and to the deviation from
KNO shape) but also to the increase of the average transverse momentum
$\langle p_{\perp}\rangle$ in the regime of ${p_{\perp}\over\sqrt{s}} 
\ll 1$.  Indeed, the $p_{\perp}$ dependence of $n$ Pomeron exchange 
amplitude is given by the expression exp $\left( -{\lambda_n p_{\perp}^2
\over n}\right)$, where $\lambda_n = R^2+\alpha'(0)\xi_n$.  Experimentally, 
for $p\bar p$-collisions, $\alpha'(0)= 0.25$ GeV$^{-2}$ \cite{ABK83}.  That
gives for the ratio 
$${\langle p_{\perp}\rangle_2 \over \langle p_{\perp}\rangle_1}
\approx \sqrt{2} \sqrt{{R^2+\alpha'\xi_1\over R^2+\alpha' \xi_2}}
\approx \sqrt{2}.$$
This ratio is in good correspondence to the experimental data
\cite{E735}.  The further increase of the $\langle p_{\perp}\rangle$
can be expected at the effective opening of the triple parton
collision, which as we see from Fig. 1 should be clearly displayed at
the LHC energy range.  An experimentally established increase of
$\langle p_{\perp}\rangle$ with energy is connected, of course, with
the gradual increase of the relative role of the double Pomeron
exchange (or double parton collisions) with energy.

There are recent results from the CDF collaboration \cite{CDF} for the
process---independent and free from theoretical schemes cross section of
double parton collision in the 3 jets data:
$$\sigma_{dp, {\rm eff}} = (14.5 \pm 1.7 {+1.7 \atop -2.3}) \; {\rm mb}.$$
The AFS experiment \cite{AFS} ($\sqrt{s}=63$ GeV) gives for the same
quantity $\sim$ 5 mb, while UA2 presents a lower limit of
$\sigma_{dp,{\rm eff}} > 8.3$ mb \cite{UA2}.

\section{Analysis of the Multiplicity Distribution Data}

Here we analyze the data on the multiplicity distributions obtained at
the energy range $\sqrt{s}$ from 30 to 1800 GeV including ISR, UA5 and
Tevatron (E735 experiment) data.  We use as a basis of our analysis
the fact that KNO scaling is well satisfied in the experimental data
through the ISR energies.  The deviation from the simple KNO scaling
at higher energies is due to another process which is incoherently
superimposed on the KNO producing process.  In Fig. 3, the differential
cross sections are displayed for the UA5 and the E735 data.  Note that
the E735 data is likely to be less reliable at low multiplicities than
the UA5 data whereas the E735 data is more statistically reliable at
the higher multiplicities.

In Fig. 3 the collider data from different energies is superimposed on
one plot.  The data have been normalized to the same value of the
variable $x={N\over \langle N_1\rangle}$ at the $x_{\rm max}$, maximum 
value ${1\over\sigma^{\rm NSD}_{in}}{d\sigma\over dx}$.  It turns out 
that the cross section at the maximum is essentially independent of energy.  
From the measurement of KNO curve at the ISR energies we know that 
$N_{\rm max} = 0.8 \langle N_1\rangle$.  Thus, we can estimate the value 
of $\langle N_1\rangle$, the average multiplicity for the processes with 
pure KNO scaling from the known value of $N_{\rm max}$ at which the cross 
section is a maximum for higher energies.  We made a least square fits
to the value of $N_{\rm max}$ and $\langle N_1\rangle$ over range
$\sqrt{s}$ from 30 GeV to 1800 GeV for the former and $\sqrt{s}$ from
30 GeV to 200 GeV (where KNO scaling is well satisfied) for the
latter.  These two quantities are well fitted by the form $a+b
\ln\sqrt{s}$ with the same coefficients over their regions of
overlap.  As we remarked above, the deviation from KNO scaling is due
to another process which is incoherently superimposed on top of the
KNO producing process.  By subtracting the KNO distribution from the
observed data we determine the shape of the competing process as shown
in Fig. 4.

The shapes of the multiplicity distribution thus found are rather
different from the KNO distrubition and is not a simple convolution of
the KNO distribution.  There must be a correlation between the
multiplicities in the two collisions.  The main characteristics of the
derived distribution is that the most probable value of the
distribution occurs at $x=2$ (or at twice the multiplicity
corresponding to the initial low energy KNO distribution).  The width
of the distribution is close to $\sqrt{2}$ times the width of that KNO
shape.  This is in quite good agreement with the picture we presented
above, which is based on the adding of the double collision of partons
of the colliding hadrons by the exchange of the two Pomerons (or two
pairs of strings) to the single parton collision which proceeds by the
exchange of one {\it effective} Pomeron.

Anyway, independently of the concrete realization of the interaction
between ultrarelativistic parton pairs, we can interpret the
population of the secondaries with maximum at $2\langle N_1\rangle$ as
a result of two independent parton-parton collisions occuring in the
same encounter.  We denote the inelastic (non-single diffractive)
cross section for this process by $\sigma_2$ whereas $\sigma_1$
denoted the inelastic cross section due to the single parton-parton
collision.  This $\sigma_1$ is characterized by the KNO scaling and,
as it follows from our analysis, it is nearly energy independent for
$\sqrt{s} \ge 200$ GeV and has a value of $(34 \pm 2)$ mb (Fig. 5).

Some remarks on the experimental data of E735 and UA5 are necessary.
In Figs. 3-5 we do not show error bars on all the points in the derived 
distributions.  The statistical errors in E735 are relatively small.  
However, the systematic error in both experiments might be sizeable.  
In E735 as well as UA5 the multiplicities in a restricted range of 
rapidity are extended to the full range by computer simulation.  There 
are corrections for secondary interactions in individual events.  The
derived distributions in Fig. 3 suffer in accuracy from the fact that
one uses subtraction of two different distributions which each have
uncertainties.  When the experimental data and the KNO distribution
are close in value then the error is greater.  Fortunately, the KNO
distribution is only the order of 10\% of the experimental data at $N
= 2\langle N_1\rangle\; (x=2)$ and consequently, the position of the
peak in the derived distribution is not strongly affected by errors
produced by taking the differences.  The whole distribution might move
up and down as a result of errors in either of the KNO or experimental
distrubitions.

The values of the cross sections represented by the curves shown in
Figs. 3,4 are given in Fig. 5.

It is seen there that, in contrast of $\sigma_1,\;\sigma_2$ is
increased with $\sqrt{s}$ and equals 16 mb at 1800 GeV.  We already
have mentioned that CDF collaboration gives the effective
double-parton collision cross section equal ($14.5 \pm 1.7 {+1.7 \atop
-2.3})$ mb.  One has to realize that in the sample of events with
double parton collisions CDF has removed events with possible triple
collisions.

We can conclude that the double (and possibly, triple) collisions
account for a large fraction of the increase in the total $p\bar
p$-cross section and, definitely, are responsible for the increase of
the $p\bar p$ inelastic cross section.

As the collision energies are increased to LHC values, it seems likely
that double and possibly triple collisions will constitute a larger
fraction of the inelastic cross section as is seen in our effective
analyses from the previous sections (see Fig. 1).

Another indication of the increasing importance of double collisions
comes from an examination of the $\langle p_{\perp}\rangle$ of the
charged particles at high multiplicity events.  As we mentioned above
there is an increase of $\langle p_{\perp}\rangle$ as multiplicity 
increases.  We note that $\langle p_{\perp}\rangle$ increases by close 
to a factor $\sqrt{2}$ from low to high multiplicity.  The picture with 
the effective double Pomeron exchange in the double-parton collision 
leads just to this $\sqrt{2}$ extra factor.  The increse of the
fraction of the multi-collisions with energy will, in general, lead to 
this well known $\langle p_{\perp}\rangle$-multiplicity relation.  We 
expect that at LHC
energies $\langle p_{\perp}\rangle$ will undergo the additional
increase.

\section{Conclusions}

In this paper we have attempted to isolate and study the double-parton 
collision mechanism from the analysis of high energy multiplicity data.  
The main result of our analysis of these data is that the non-single
diffractive inelastic cross section consists of two components.  The
first component corresponding to the single parton collisions is
characterized by being practically independent of $\sqrt{s}$ cross
section for $\sqrt{s} > 200$ GeV, whereas the second part of the
$\sigma_{in}^{NSD}$ increases significally with energy and achieves the
value 16 mb at $\sqrt{s} = 1.8$ TeV.  That part was attributed here to
the inclusion in the collision process of the double (and, maybe,
triple) parton collisions.  Thus, the increase of the
$\sigma_{in}^{NSD}$ at high $\sqrt{s}$ is almost entirely due to the
multi-parton collisions.

On the other hand, we know from the experiment at lower energies where
the single parton collisions indeed dominate that $\sigma_{in}^{NSD}$
is increasing with $\sqrt{s}$.  This indicates that at higher energies
$\sqrt{s} > 200$ GeV the inelastic cross section due to the single
collision goes to the saturation, whereas the double collisions give
the increasing part of the $\sigma_{in}^{NSD}$.  One can go further
and conjecture that the same saturation will occur for the part of
$\sigma_{in}$ connected with the double parton collisions at much
higher energies whereas the $\sigma_{in}$ due to the triple collisions
will still increase with $\sqrt{s}$, etc., until asymptotically the
total inelastic cross section (without the diffractive part) will set
a constant value $\sigma_{in} = \sigma_1^{\rm (sat)} + \sigma_2^{(\rm
sat)} + \ldots $.  Of course, we cannot say anything about the behavior 
of $\sigma_{\rm el}$ and $\sigma_{\rm dif}$ at this hypothetical limit.

We need to emphasize that in this paper we have often used the notion
of the ``soft'' Pomeron exchange in the process of parton collisions.
Of course, this notion should be understood as an effective tool of
the description of the ultrarelativistic parton collision and have
played only the illustrative role and the trend in our analysis.
However, the picture of the multi-parton collisions described here and
the corresponding values for $\sigma_1$ and $\sigma_{2+3}$ extracted
from our above analysis of the multiplicity are independent of the
concrete mechanism of the parton-parton interaction. 

\acknowledgements

We are grateful to Berndt M\"uller for interesting discussions and for
reading the manuscript.  S.M. acknowledges the useful discussions with
Leonid Frankfurt and Eugene Levin.  W.W. acknowledges the useful
discussion with Francis Halzen.  We are pleased to acknowledge the 
help of Glenn Doki and Dan Cronin-Hennessy in the numerical calculations.
This work is supported in part by a Department of Energy Grant No.
DE-FG02-96ER-40945.

\begin{figure}
\caption{ a) Topological cross sections $\sigma_N$ in the quasi-eikonal
approximation with exchanges of three effective ``soft'' Pomerons for
$\sqrt{s}$ = 546, 900, 1800 and $14.10^3$ GeV. b) Topological cross
sections resulting from double and triple parton collisions for
$\sqrt{s}$ = 546, 900, 1800 and $14.10^3$ GeV.}
\label{fig1}
\end{figure}

\begin{figure}
\caption{Same in Fig. 1 for $\sqrt{s}$ = 100, 200 and 300 GeV.}
\label{fig2}
\end{figure}

\begin{figure}
\caption{A comparison of multiplicity distributions at different
values $\sqrt{s}$.  The distributions are normalized at the maximum
value of ${d\sigma\over dx}$ where $x={n\over \langle n_1\rangle}$.
The solid curve is the KNO distribution from ISR data.}
\label{fig3}
\end{figure}

\begin{figure}
\caption{Multiplicity distributions obtained by taking the difference
between the $p\bar p$ collider data and the KNO distribution.}
\label{fig4}
\end{figure}

\begin{figure}
\caption{Cross sections for the single $(\sigma_1)$ and multiparton
(double and triple) $(\sigma_2)$ parton collisions as a function of
$\sqrt{s}$.}
\label{fig5}
\end{figure}

\vfill
\eject

\section*{Appendix A}

According to our ideology, at high energies $(\sqrt{s} \gg 1$ GeV) the
single parton collisions which may be described by the effective single 
Pomeron exchange and which are characterized by the scaled multiplicity
distribution, must be applied only up to the energy range where
KNO scaling takes place.  That means that the average multiplicity of 
the hadrons (in $pp$- and $\bar pp$- interactions) for the single 
collisions $\langle N_1\rangle$ must be obtained {\it only}
from this energy range.  The usually used total range of the existing
energies up to highest ones (Tevatron), being important, of course,
however, from the developed point of view, disguises the mechanisms of
the high energy collisions.  Thus, we describe, in the spirit of the
Regge picture, the average multiplicity $\langle N_1\rangle$ as $a+b
\ln{s\over s_0}$ and find the coefficients $a$ and $b$ from the data
of the high energy $pp$-collisions in the range of $\sqrt{s}\;(11.5 \div
62,6)$ GeV \cite{16}\footnote{When our paper was written we became
aware of the recent similar treatment of the average multiplicity
$\langle N_1\rangle$ as a linear logarithmic function for ``soft''
component \cite{17}.}
As a result of the fit, we have for the coefficients $a$ and $b$:
$a= -3.874$, $b = 2.099$ with $\chi^2 =7.085/7$.

We use the corresponding $\langle N_1\rangle$ as input in calculation
of the multiplicity distributions which takes into account the
multiparton $(n=2,3)$ collisions (see Appendix B).

\section*{Appendix B}

We briefly describe here the results for the cross section of the
production of $N$ particle in the quasi-eikonal approximation of the
Regge approach (or, equivalently, as we emphasized in the text, in the
model of the gluonic strings \cite{PEV76,ABK83}).  These results are
represented graphically in Fig. 1.  

We are interested in the inelastic cross-section $\sigma_{in}(N,s)$ not 
including the diffraction processes.  The ignorance of the ``enhanced'' 
graphs allows us to consider that the long distant correlations among 
particles belonging to one shower-ladder are absent \cite{KAM73}.  
Similarly, we suppose that the partons from different showers are 
uncorrelated also.  This permits us to assume that the distributions 
$P_n(N)$ of the $N$ produced particle (in full phase space) in the $n$ 
shower events is the Poisson type:\footnote{Use of the concrete form
of (1) is not crucial for obtaining the relation (5).  It has a place
for normalized $P_n(N)$ with $\langle N_n\rangle = n \langle
N_1\rangle$  and large $\langle N_1\rangle$.  The relation (4), of course, 
does not depend on (1).  Figs. 1-2 show only the trends.}

\begin{equation}
P_n(N) = {\langle N_n\rangle^N\over N!}\; e^{-\langle N_n\rangle}
\label{A1}
\end{equation}
\begin{equation}
\langle N_n\rangle = n\langle N_1\rangle = n\left(a+b \ln{s\over
s_0}\right).
\end{equation}

We can write for $\sigma_{in}(N,s)$
\begin{equation}
\sigma_{in}(N,s) = \sigma_1P_1(N) + \sigma_2(P_2(N) +
\sigma_3P_3(N) + \ldots , \label{A2}
\end{equation}
where $\sigma_n(\xi_n)$ is the cross section for the production of $n$
ladders \cite{ABK83} and $\xi_n = \ln\left( {s\over s_0n^2}\right)$.
For $\sigma_n (\xi_n)$ we have \cite{ABK83}
\begin{equation}
\sigma_n(\xi_n) = {\sigma_p\over nZ_n} \left( 1-e^{-Z_n}
\sum_{k=0}^{n-1} {Z_n^k\over k!} \right) \label{A3}
\end{equation}
with
$$
\sigma_p = 8\pi\gamma \left( {s\over s_0}\right)^{\Delta},\quad Z_n =
{2C\gamma\over R^2 + \alpha'_p\xi_n} \left({s\over s_0
n^2}\right)^{\Delta}.$$

The parameters in these expressions are fixed by the fits of the
experimental data on
$\sigma_{\rm tot}$ and ${d\sigma_{\rm el}\over dt}$ for $pp$- and 
$\bar p p$-collisions \cite{PEV76}:

$$\gamma_p = 3.64\;({\rm GeV})^{-2}, \quad R^2 = 3.56\;({\rm GeV})^{-2},
\quad C=1.5,$$

$$\Delta = \alpha_p(0) -1 = 0.08, \quad \alpha'_p = 0.25 ({\rm
GeV})^{-2} (c=1).$$
We retain three terms in (\ref{A2}) since the quadruple (or
four-parton pair) collision has small effect (see Fig. 1 which shows
that even the triple collisions appear noticable only at the energe
range of LHC).

Summing (\ref{A2}) over $N$ from 1 to $\infty$ with $\langle
N_n\rangle \gg 1$ we obtain for the total inelastic cross section
\begin{equation}
\sigma_{in}(s) = \sigma_1+\sigma_2+\sigma_3 + O(\sigma_4)
\label{A4}
\end{equation}

Fig. 1a) shows the $N$-distributions of $\sigma_{in}(N,s)$ at
$\sqrt{s}= 0.55$; 0.9; 1.8 and 14 TeV.  At lower $\sqrt{s}$ the second
peak of $\sigma(N,s)$ is not resolved.  Fig. 1b) shows the behavior of
the parts of $\sigma(N,s)$ which can be attributed to the double (and
triple) collisions.

In the table we present values of the inelastic ``partial'' cross
sections $\sigma_1,\; \sigma_2$ and $\sigma_3$ corresponding to the
single, double and triple parton collisions in the Regge quasi-eikonal
approach.

We see that in this approach all $\sigma_i\; (i=1,2,3)$ increase with
$\sqrt{s}$ whereas our analysis in the text above indicates that the
``single'' collision contribution $\sigma_1$ extracted from the 
experimental data on the multiplicity distribution practically is 
independent of $\sqrt{s}$.

Furthermore, $\sigma_1+\sigma_2+\sigma_3$ are systematically lower
than the experimental cross section $\sigma_{in}^{\rm tot}(s)$ for the
non-single diffraction events, whereas the corresponding theoretical
values of this cross section resulting from the summation of {\it all}
quasi-eikonal pomeron graphs \cite{PEV76,ABK83}
$$\sigma_{in}^{\rm tot}(s) = \sigma_pf(Z)$$
where
$$f(Z) = \sum_{v=1}^{\infty} {(-Z)^{v-1}\over vv!} = {1\over Z}
\int_0^Z (1-e^{-x}) {dx\over x} = {1\over Z} \left(\Gamma(0,Z) + \ln 
(\gamma_EZ)\right),$$
with $\gamma_E = 1.78 \ldots$.  Euler constant and $\Gamma(\alpha,Z)$-
incomplete gamma function, are in excellent agreement with experimental
data for $\sigma_{in}^{\rm tot}$.

Fig. 2a and 2b show for completeness the $\sigma(N,s)$ for lower
$\sqrt{s}$ (but higher than 62 GeV) ($\sqrt{s} =$ 100, 200, 300 GeV).
On these figures at $\sqrt{s}= 200$ GeV the shoulder, corresponding to
the double collision is clearly seen.

\begin{table}
\begin{tabular}{|c|r|r|r|c|}
$\sqrt{s}$, TeV &$\sigma_1$, mb &$\sigma_2$, mb &$\sigma_3$, mb
&$\sigma_1+\sigma_2+\sigma_3$, mb \\
\hline \\
0.55 &21.94 &9.57 &5.25 &36.76 \\
0.9 &22.72 &10.16 &5.71 &38.59 \\
1.8 &23.84 &10.72 &6.34 &40.90 \\
14 &27.19 &14.70 &8.43 &50.32
 
\end{tabular}
\end{table}

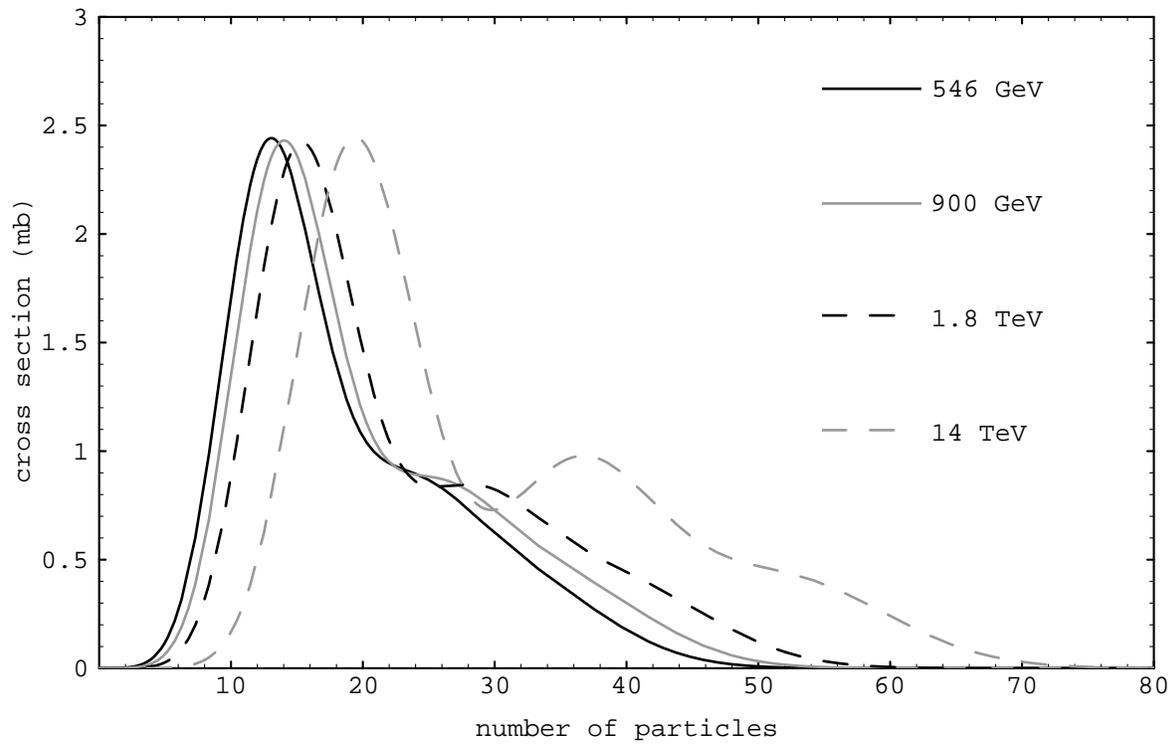
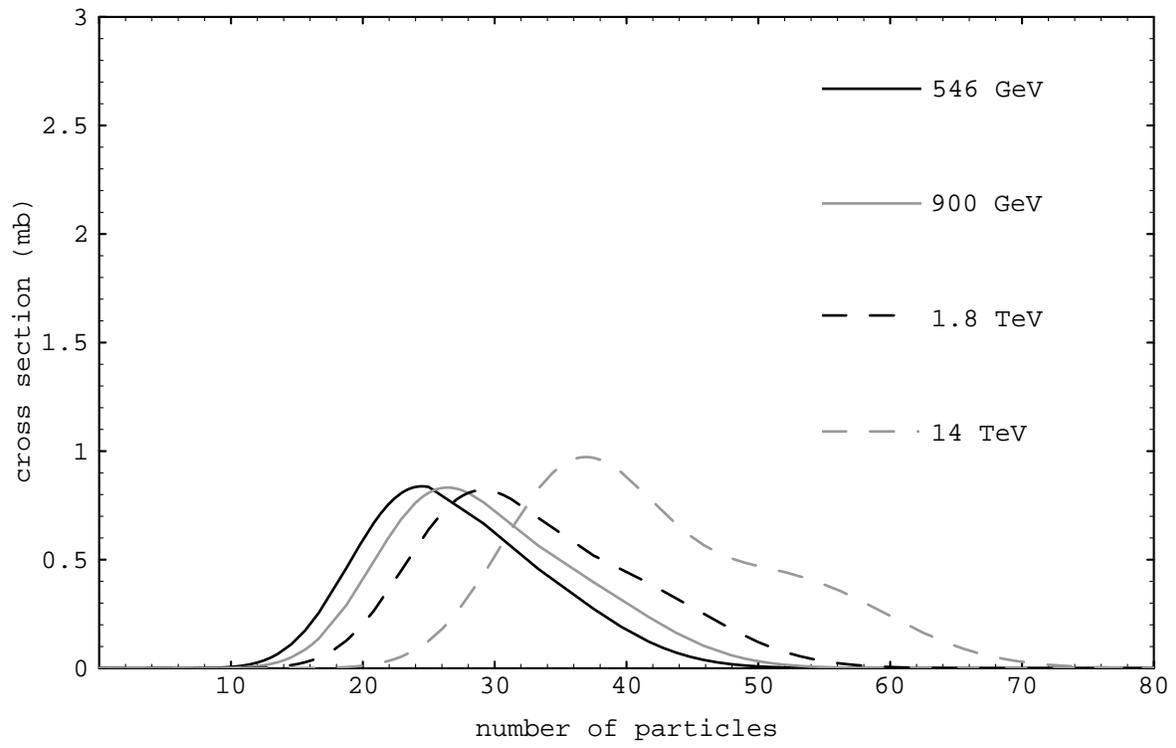
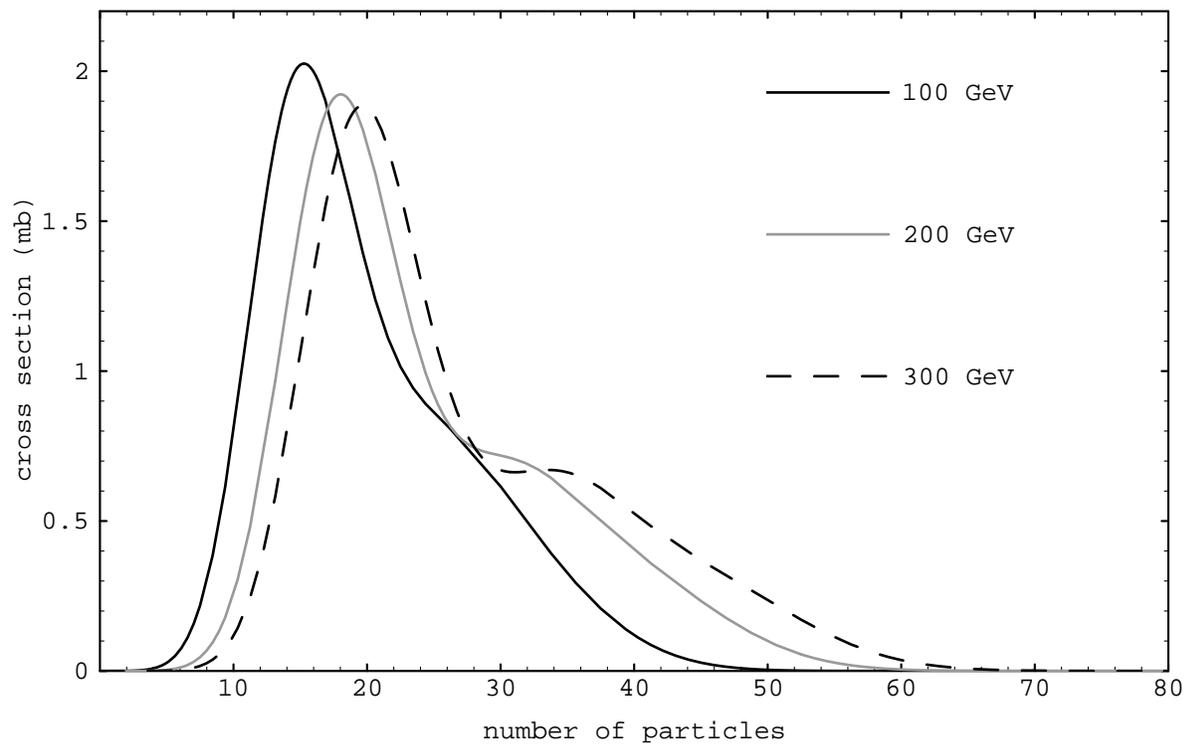
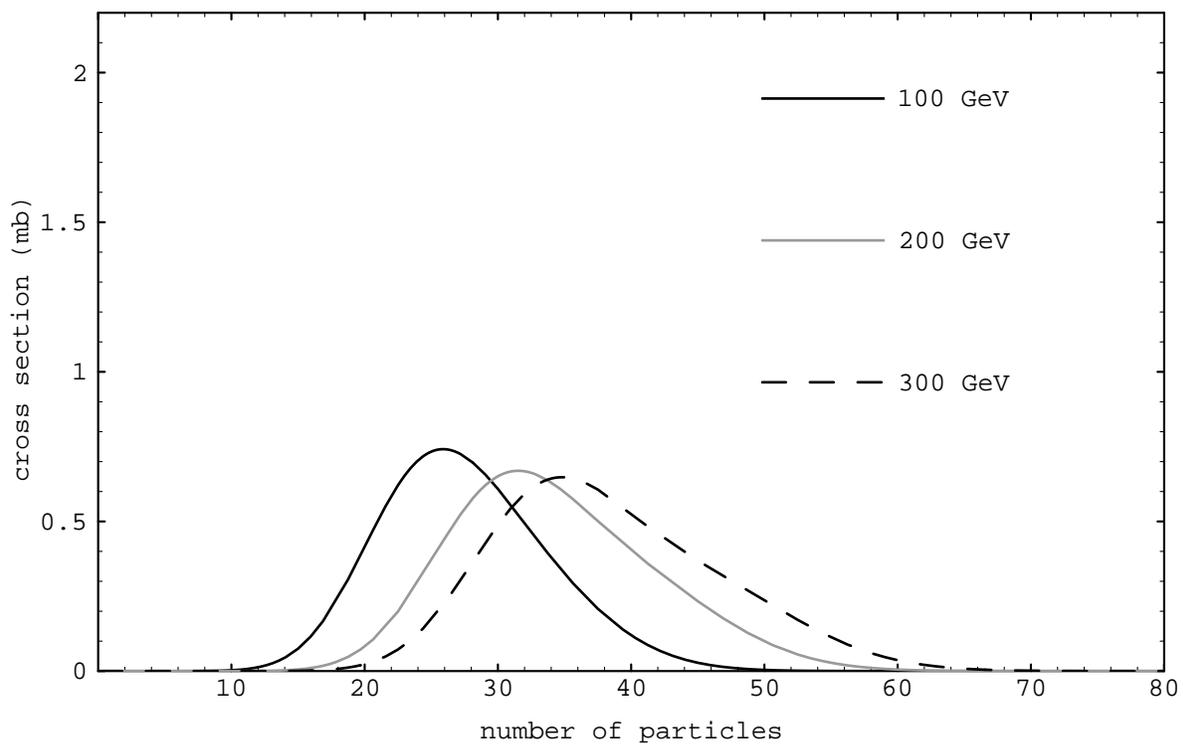
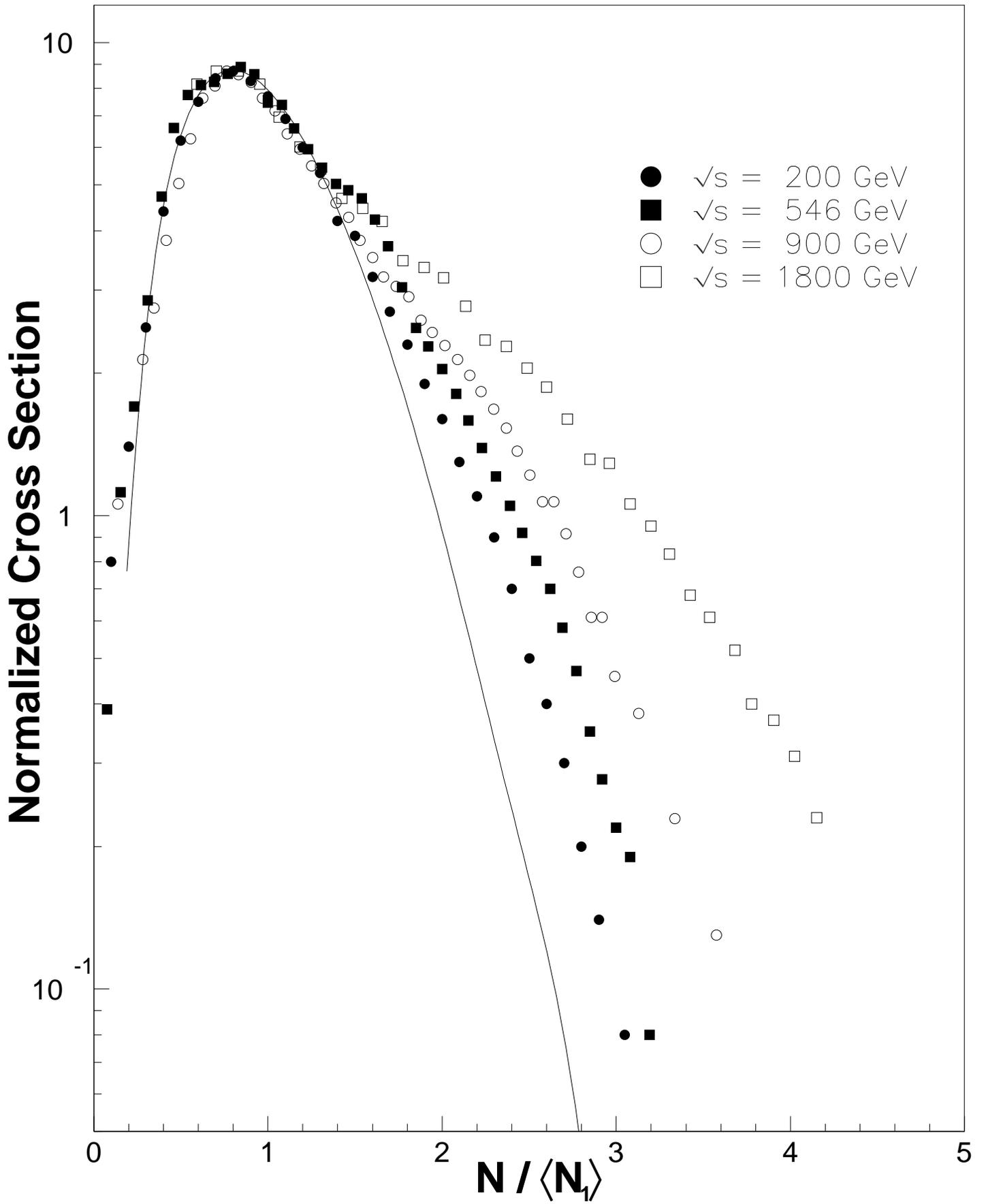
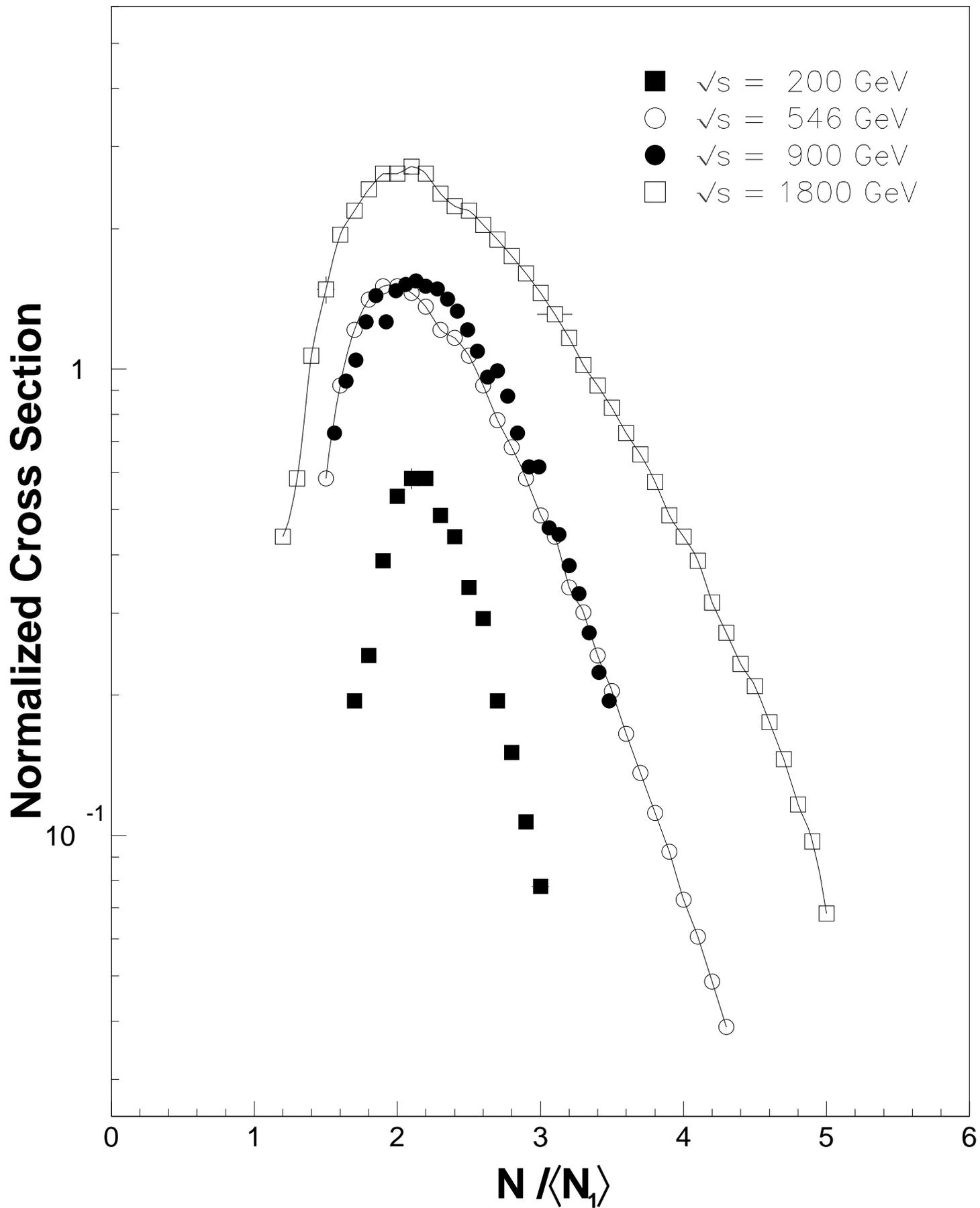
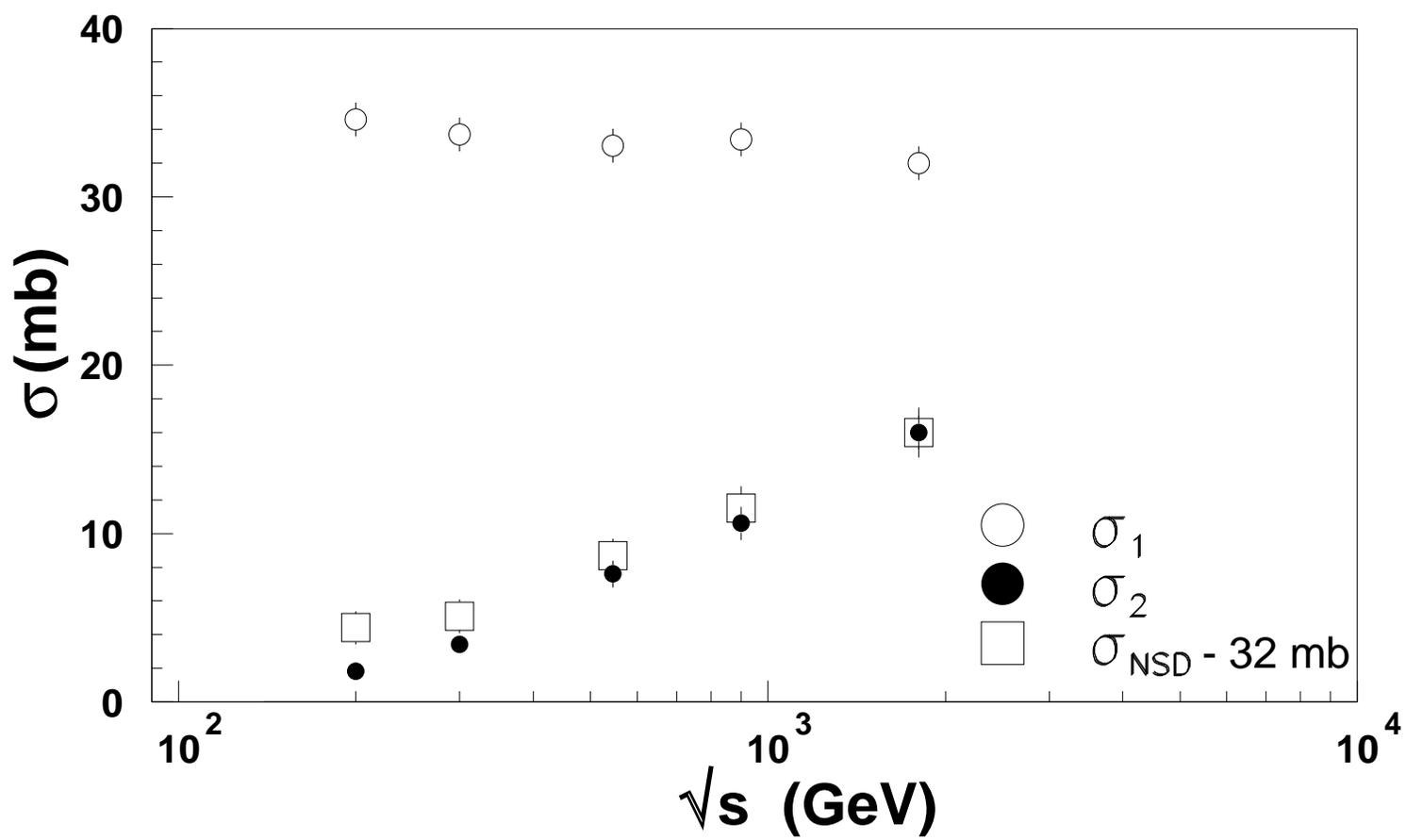
\end{document}